\documentclass[twocolumn]{IEEEtran}

\usepackage{graphicx,amsmath,epsfig,amssymb,verbatim,cite,amsopn}

\newcommand{\pictwidth}{90truemm}
\newcommand{\sa}{\sigma_a}

\newcommand{\sx}{\Omega_x}

\newcommand{\sn}{\sigma_n}
\newcommand{\hq}{q}
\newcommand{\hr}{v}
\newcommand{\hw}{\omega}

\newcommand{\gap}{G}
\newcommand{\deltaa}{L}
\newcommand{\power}{\xi}
\newcommand{\Ei}{\mathrm{Ei}}

\newcommand{\Teleng}{\small{Department of Information Engineering}}
\newcommand{\RSISE}{Research School of Information Sciences and Engineering}
\newcommand{\ANU}{The Australian National University, ACT 0200, AUSTRALIA}
\newcommand{\AuthorOne}{\small{Rasika R. Perera}}
\newcommand{\AuthorTwo}{Tony S. Pollock*}
\newcommand{\AuthorThree}{Thushara D. Abhayapala*\thanks{
                          T. S. Pollock and T. D. Abhayapala have appointments with National ICT Australia (NICTA).
                          Part of this work has been presented at the Australian Communications Theory Workshop (AUSCTW), 2005.}}

\title{Performance of Gaussian Signalling in Non Coherent Rayleigh Fading Channels}
\author{\authorblockN{\textit{\AuthorOne, \AuthorTwo,\ and
\AuthorThree}}\\
\authorblockA{\Teleng\\\RSISE\\\ANU}}

\begin{document}

\maketitle

\begin{abstract}
The mutual information of a discrete time memoryless Rayleigh
fading channel is considered, where neither the transmitter nor
the receiver has the knowledge of the channel state information
except the fading statistics. We present the mutual information of
this channel in closed form when the input distribution is complex
Gaussian, and derive a lower bound in terms of the capacity of the
corresponding non fading channel and the capacity when the perfect
channel state information is known at the receiver.
\end{abstract}


\section{INTRODUCTION}
An independent and identically distributed (iid) Gaussian is the
capacity achieving input distribution for additive white Gaussian
noise (non-fading) channel, a Rayleigh fading channel when the
Channel State Information (CSI) is perfectly known at the
receiver, and when the CSI is known to both the transmitter and
the receiver. However, when CSI is not known by neither the
transmitter nor the receiver, the capacity achieving distribution
is not Gaussian \cite{DTMRayleigh-Shami-2001}. Therefore, it is of
practical interest to find the achievable information rate of non
coherent Rayleigh fading channels when the input distribution is
complex Gaussian.

Fading channels have been studied in depth and a plethora of
literature is available on the upper and lower achievable rates
over the wireless media; refer \cite{Proakis-Fading-Ch-1998},
\cite{Shamai-csi-some-ch-1999} for a summary. However, most of the
results were presented under various channel models applying
constraints for mathematical representations, and the availability
of (CSI) at the transmitter and receiver.

The capacity of fading channels when the CSI is perfectly known at
the receiver was investigated initially by Ericson
\cite{Ericson-1970}, later by Lee \cite{W-C-Y1990}, and Ozarow,
Shamai and Wyner \cite{Shamai-Osarow-Wyner-1994}. This capacity is
calculated in an average sense due to the time varying nature of
the signal to noise ratio (SNR). The fading channel with CSI at
the receiver alone and at both the transmitter and the receiver
was extensively studied in \cite{Goldsmith-Varaiya-csi-1997},
\cite{Goldsmith-Mohamed-csi-2000}.

The iid Rayleigh fading channel with no CSI was studied by Faycal
\cite{DTMRayleigh-Shami-2001}, \cite{Abou-1996}, where it was
shown that the capacity achieving input distribution is discrete
with finite number of mass points with new emerging points as SNR
increases. These mass point distribution tends to be uniform as
SNR approaches infinity, deviating much form that of a Gaussian.
The non coherent time selective Rayleigh fading channel has been
further investigated by Yingbin and Venugopal
\cite{Yingbin-nocsi-corre-2004} and derived upper and lower bounds
on the capacity at high SNR.

In this paper, we determine how the Gaussian input distribution
can contribute in non coherent Rayleigh fading channel. We achieve
this by expressing the mutual information in closed form using
\textit{Gauss-Hermit Quadrature}\footnote{One of the best methods
that can be used to evaluate the integrals of the type
$\int_{0}^{\infty}{e^{-x^2}}g(x){dx}\simeq\,\sum_{i=1}^{m}{\omega_{i}}{g(x_i)}$
in terms of proper weights $\omega_{i}$ and the roots $x_{i}$ of
Hermit Polynomials $H_{m}(x)$.} with a simple lower bound on it
and subsequently identifying the maximum deviation of the actual
capacity achieved with a discrete input in the presence of
Gaussian input.



\section{SYSTEM MODEL}
Consider the Rayleigh fading channel,
\begin{equation}\label{ray-channel}
    y={a}{x}+{n}
\end{equation}
where $y$ is the complex channel output, $x$ is the complex
channel input, $a$ and $n$ represent the fading and noise
components associated with the channel. It is assumed that $a$ and
$n$ are independent zero mean circular complex Gaussian random
variables. Also assume that ${\sa^2}/{2}$ and ${\sn^2}/{2}$ are
the equal variance of real and imaginary parts of the complex
variables $a$ and $n$ respectively and the time index in
\eqref{ray-channel} is omitted for simplicity. The random
variables $a$, $x$, and $n$ are considered to be independent of
each other. The input $x$ $\epsilon$ $X$ is average power limited:
$E[|x|^2]=\sx^2\leq P$. The constant
$\gamma=-\int_{0}^{\infty}{e^{-y}}{\log{y}}{dy}\approx0.5772...,$\,
denotes the Euler's constant. All the differential entropies and
the mutual information are defined to the base ``e'', and the
results are expressed in ``nats''. It is assumed that neither the
receiver nor the transmitter has the knowledge of channel state
information other than the statistics.

\section{THE MUTUAL INFORMATION}

The Mutual information between the input and output of a Rayleigh
fading channel can be expressed as \cite{Taricco-1997}
\begin{align}
    I(X;Y)&=\int_{0}^{\infty}{\int_{0}^{\infty}}{p_{Y|X}(y|x)}{p_{X}(x)}\nonumber\\
    &{\times}\,\,{\log\left[{\frac{p_{Y|X}(y|x)}{\int_{0}^{\infty}{p_{Y|V}(y|v)}{p_{V}(v)}{dv}}}\right]}\,{dx}{dy}
\label{mutual-information}
\end{align} considering the probability distribution of the
magnitudes of the input and output random variables $X$ and $Y$.
It should be noted here that since we only consider the
distribution of magnitudes of the random variables, the integral
in \eqref{mutual-information} is taken from $0$ to ${\infty}$. The
conditional probability density function (pdf) $p_{Y|X}(y|x)$
\cite{Abou-1996}\cite{Taricco-1997} is given by
\begin{equation}\label{gen-cond-output-prob}
    p_{Y|X}(y|x)=\frac{2{|y|}}{{\sn^2}+{\sa^2}{|x|^2}}
    {\exp\left({\frac{-|y|^2}{{\sn^2}+{\sa^2}{|x|^2}}}\right)}.
\end{equation} Assume the average mean squared power of
both fading $A$, ($a$ $\epsilon$ $A$) and noise $N$, ($n$
$\epsilon$ $N$) are unity. This assumption is valid since the
effective received power at the receiver is the combination of
both $\sa^2$ and $\sx^2$ and the SNR is the ratio between the
average received power and the average noise power. Therefore, the
same output exists for various $\sa^2$ and $\sn^2$ on the
appropriate selection of $\sx^2$. With this assumption,
\eqref{gen-cond-output-prob} can be written as
\begin{equation}\label{cond-output-prob}
    p_{Y|X}(y|x)=\frac{2{y}}{1+x^2}{\exp\left({\frac{-y^2}{1+x^2}}\right)}.
\end{equation}Without loss of generality, the magnitude sign is removed in
\eqref{cond-output-prob} and the same notation will be used
throughout the rest of this paper.

The mutual information \cite{T-M-Cover-1991} in
\eqref{mutual-information} can be simplified to
\begin{subequations}
\begin{align}\label{mutual-information-simplified1}
I(X;Y)&=h(Y)-h(Y|X)\\\label{mutual-information-simplified2}
&=-\int_{0}^{\infty}{p_{Y}(y){\log{p_{Y}(y)}}}{dy}\nonumber\\
&-\frac{1}{2}{\int_{0}^{\infty}
{p_{X}(x)}{\log(1+x^2)}{dx}}+{\log{2}}-({1+\frac{\gamma}{2}})
\end{align}
\end{subequations} where the first term in
\eqref{mutual-information-simplified2} is the channel output
entropy $h(Y)$. This was originally proven by Taricco
\cite{Taricco-1997} deriving an analytical expression for the
channel capacity using Lagrange optimization method with an
additional constraint.

\section{GAUSSIAN INPUT IN NON COHERENT RAYLEIGH FADING}
Recall the channel model \eqref{ray-channel}, and assume the input
distribution is Gaussian. Then the distribution of both the real
and imaginary parts of $x$ are independent and Gaussian.
Therefore, the distribution of the $|x|$ is Rayleigh with the pdf
\cite{A-B-Carlson-1986}
\begin{equation}\label{Ray_PDF}
    p_{X}{(x)}=
            {\frac{2{x}}{\sx^2}}{\exp\left({\frac{-x^2}{\sx^2}}\right)}, \,x\geq0.
\end{equation}
It is assumed that both the real and imaginary parts of input have
equal variance ${\sx^2}/{2}$. The magnitude sign is omitted in
\eqref{Ray_PDF} as mentioned in the previous section.

\subsection{Output Conditional Entropy} Having described the
input distribution $p_X(x)$ for non coherent Gaussian input
channel, we now focus on the output conditional entropy $h(Y|X)$
in \eqref{mutual-information-simplified2}. By substituting
\eqref{Ray_PDF} in \eqref{mutual-information-simplified2} (except
the first term described as $h(Y)$), we have
\begin{align}
    h(Y|X)&={\int_{0}^{\infty}\left[
    {\frac{{x}}{\sx^2}{\exp\left({\frac{-x^2}{\sx^2}}\right)}}{\log(1+x^2)}\right]{dx}}\nonumber\\
    &-{\log{2}}+({1+\frac{\gamma}{2}}).
\label{hyx-one}
\end{align}
With the detailed proof provided in Appendix A, we can reduce
\eqref{hyx-one} to
\begin{equation}\label{hyx-final}
    h(Y|X)={-}\frac{1}{2}{\exp\left({\frac{1}{\sx^2}}\right)}{\Ei\left(\frac{-1}{\sx^2}\right)}
           -{\log{2}}+({1+\frac{\gamma}{2}}),
\end{equation}
where the exponential integral
$\Ei(x)={-}{\int_{-x}^{\infty}}{{e^{-t}}/{t}}\,{dt}$. Note that
the channel capacity when the CSI  is perfectly known at the
receiver is \cite{Proakis-Fading-Ch-1998}, \cite{Ericson-1970},
\cite{W-C-Y1990},
\begin{equation}\label{cap-withcsi}
    C_{\text{rcsi}}=-{\exp\left({\frac{1}{snr}}\right)}{\Ei\left(\frac{-1}{snr}\right)},
\end{equation}where $snr=\sx^2$ since $\sn^2=1$.
Therefore, $h(Y|X)$ in non coherent Rayleigh fading with Gaussian
input can be expressed as
\begin{equation}\label{hyx-final2}
    h(Y|X)=\frac{1}{2}{C_{\text{rcsi}}}-{\log{2}}+({1+\frac{\gamma}{2}}).
\end{equation}

\subsection{Output Entropy} The output pdf
$p_Y(y)=\int_{0}^{\infty}{p_X(x)}{p_{Y|X}(y|x)}{dx}$ for the
Gaussian input can be written as
\begin{equation}\label{output-pdf}
    p_Y(y)=\int_{0}^{\infty}{\frac{{2}{x}}{\sx^2}}
    {\exp\left({\frac{-x^2}{\sx^2}}\right)}\frac{2{y}}{1+x^2}{\exp\left({\frac{-y^2}{1+x^2}}\right)}{dx}.
\end{equation}
Substituting \eqref{output-pdf} in the first term of
\eqref{mutual-information-simplified2} gives
\begin{align}
    h(Y)&
    =-\int_{0}^{\infty}\int_{0}^{\infty}{\frac{{4}{x}{y}}{{\sx^2}{(1+x^2)}}}
{\exp\left({-}{\frac{x^2}{\sx^2}}-{\frac{y^2}{1+x^2}}\right)}{dx}\nonumber\\
&{\times}\,{\log{\left[\int_{0}^{\infty}{\frac{{4}{x}{y}}{{\sx^2}(1+x^2)}}
{\exp\left({-}{\frac{x^2}{\sx^2}}-{\frac{y^2}{1+x^2}}\right)}{dx}\right]}}{dy}.
\label{out-entro}
\end{align}

To the best of our knowledge, this integral can not be evaluated
analytically $\forall\,\,\sx^2$. In the following section we show
the use of Gauss-Hermit polynomials to drive a closed form
expression.

\subsection{Gaussian Quadrature and Hermit Polynomials}
A common method for approximating a definite integral is
$\int_{a}^{b}{\omega(x)}{f(x)}{d{x}}\simeq\sum_{i=1}^{\hq}{A_{i}{f(x_i)}}$,
which is called Gauss-quadrature assuming the moments are defined
and finite or bounded of the function ${\omega(x)}$
\cite{Gaussian-quadrature-formulas-stroud}. The Gaussian
quadrature formula has a degree of precision or exactness $m$ if
the solution is exact whenever $f(x)$ is a polynomial of degree
$\leq m$ or equivalently, whenever $f(x)=\{1,x,....,x^m\}$ and it
is not exact for $f(x)=x^{m+1}$. The $x_{i}$ are called the nodes
of the formula and $A_i$ are called coefficients (or weights). If
$\omega(x)$ is non negative in $[a,b]$, then $n$ points and
coefficients can be found to make the solution exact for all
polynomials of degree $\leq 2{\hq}-1$ and it is the highest degree
of precision which can be obtained using $\hq$
points \cite{Gaussian-quadrature-formulas-stroud}.\\\\
If the $\omega(x)$ is in the form of $e^{-x^2}$, the solution to
the integral can be found using the roots (nodes) of the Hermit
polynomial $H_{\hq}(x)$ \cite{Gaussian-quad-zero-to-infty} and the
weights are given by
\begin{equation}\label{hermit-weights}
    \hw_{i}=\frac{{2^{\hq-1}}{\hq!}{\sqrt{\pi}}}{{\hq^2}{\left[H_{\hq-1}(x_{i})\right]^2}}.\nonumber
\end{equation} The roots and the weights are excessively given in \cite{Gaussian-quad-zero-to-infty}
for $a=0$ and $b=\infty$ with $\hq=15$.

\subsection{Output entropy $h(Y)$ in Closed form}

Define $t^2=x^2/\sx$, where $dx={\sx}{dt}$, then substitution into
\eqref{output-pdf} gives
\begin{equation}\label{py-siso-hermit1}
        p_Y(y)=\int_{t=0}^{t=\infty}{e^{-t^2}{\frac{{2}{t}{y}}
        {(1+{\sx}{x^2})}{\exp\left[{\frac{-y^2}{2(1+{\sx}{x^2})}}\right]}{dt}}}.
\end{equation}
This integral is in the form of
$\int_{a}^{b}{\phi}(\hr){\hw}(\hr){d\hr}$ where
$\hw(\hr)\equiv{e^{-t^2}}$. Therefore it can be evaluated using
Hermit polynomials in the form of
$p_{Y}(y)=\sum_{j=1}^{\hq}{\hw_j}{f(\hr_j)}$. The quantities
$\hr_j$ and $\hw_j$ are the roots and the weights of the Hermit
polynomials respectively. Applying these weights and roots in
\eqref{py-siso-hermit1} we get
\begin{equation}\label{py-siso-hermit-final}
    p_Y(y)=\sum_{j=1}^{\hq}{\hw_{j}}{\frac{{2}{\hr_{j}}{y}}
    {(1+{\sx}{\hr_{j}^2})}}{\exp\left[{\frac{-y^2}{2(1+{\sx}{\hr_{j}^2})}}\right]}.
\end{equation}
Using this result, the output entropy $h(Y)$ can be written as
\begin{align}
    h(Y)&=-\int_{0}^{\infty}\left\{\sum_{\ell=1}^{\hq}{\omega_{\ell}}{\frac{{2}{\hr_{\ell}}{y}}
    {(1+{\sx}{\hr_{\ell}^2})}}{\exp\left[{\frac{-y^2}{2(1+{\sx}{\hr_{\ell}^2})}}\right]}\right\}\nonumber\\
    &{\times}\,{\log}{\left\{\sum_{j=1}^{\hq}{\omega_{j}}{\frac{{2}{\hr_{j}}{y}}
    {(1+{\sx}{\hr_{j}^2})}}{\exp\left[{\frac{-y^2}{2(1+{\sx}{\hr_{j}^2})}}\right]}\right\}}{dy}.
    \label{hy-siso-hermit1}
\end{align}Taking the integration inside and substituting
$t^2=y^2/[2(1+\sx\hr_{j}^2)]$ in \eqref{hy-siso-hermit1}, the
$\ell^{th}$ term where $\ell=1,....,\hq$ can be written as
\begin{align}
    h(Y)_{\ell}=&-\int_{0}^{\infty}{e^{-t^2}}({4}{\hw_{l}}{\hr_{\ell}}{t})\log\left\{
    \sum_{j=1}^{\hq}{\frac{2{\sqrt{2}}{\hw_j}{\hr_j}{t}}{(1+\sx{\hr_{j}^2})}}\right.\nonumber\\
    &\left.{\times}\,{\sqrt{(1+\sx{\hr_{\ell}^2})}}\,{\exp\left[{{-t^2}\frac{(1+\sx{\hr_{\ell}^2})}{(1+\sx{\hr_{j}^2})}}\right]}\right\}{dt}.
\label{hy-L-siso-1}
\end{align} The integral in this $l^{th}$ term
has the form of $\int_{a}^{b}{\phi}(\hr){\hw}(\hr){d\hr}$ where
$\hw(\hr)\equiv{e^{-t^2}}$. We can now simplify
\eqref{hy-L-siso-1} using Hermit polynomials as
\begin{align}
    h(Y)_{\ell}&=\sum_{i=1}^{r}{\hw_i}({4}{\hw_{\ell}}{\hr_{\ell}}{\hr_i})\log\left\{
    \sum_{j=1}^{\hq}{\frac{2{\sqrt{2}}{\hw_j}{\hr_j}{\hr_i}}{(1+\sx{\hr_{j}^2})}}\right.\nonumber\\
    &\left.{\times}\,{\sqrt{(1+\sx{\hr_{\ell}^2})}}\,{\exp\left[{{-\hr_{i}^2}\frac{(1+\sx{\hr_{\ell}^2})}{(1+\sx{\hr_{j}^2})}}\right]}\right\}.
\label{hy-L-siso-2}
\end{align}
The output entropy $h(Y)=-\sum_{\ell=1}^{\hq}{h(Y)_{\ell}}$ can be
shown as
\begin{align}
    h(Y)&=-\sum_{\ell=1}^{\hq}\sum_{i=1}^{r}({4}{\hw_i}{\hr_i}{\hw_{\ell}}{\hr_{\ell}})\log\left\{
    \sum_{j=1}^{\hq}{\frac{2{\sqrt{2}}{\hw_j}{\hr_j}{\hr_i}}{(1+\sx{\hr_{j}^2})}}\right.\nonumber\\
    &\left.{\times}\,{\sqrt{(1+\sx{\hr_{\ell}^2})}}\,{\exp\left[{{-\hr_{i}^2}\frac{(1+\sx{\hr_{\ell}^2})}{(1+\sx{\hr_{j}^2})}}\right]}\right\}.
\label{hy-siso-closed}
\end{align}
The $h(Y)$ presented in the closed form in
\eqref{hy-siso-closed} using Gauss-Hermit quadrature is very
useful in finding the mutual information for any SNR and the
computational time is much less than the numerical integrations to
be carried out with high accuracy. The mutual information can be
found subtracting \eqref{hyx-final2} from \eqref{hy-siso-closed}.
Fig. 1 depicts the mutual information obtained using the
Gauss-Hermit polynomial method with the channel capacity
\cite{DTMRayleigh-Shami-2001}.

\begin{figure}
  \centering
  \includegraphics[width=\pictwidth]{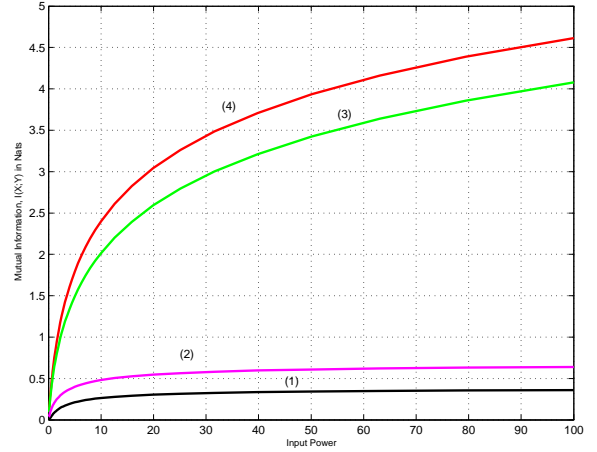}\\
  \caption{The mutual information with Gaussian input is below
  the channel capacity achieved by a discrete input. (1) The mutual information with gaussian
  input. (2) The channel capacity with a discrete input. (3) The channel capacity, $C_{\text{rcsi}}$ with CSI.
  (4) The non fading channel capacity, $C_{\text{cnf}}$.}\label{figone}
\end{figure}

Since the closed form expression obtained in the previous section
is intricate with no straightforward or easy method to attain the
result, we will show how to derive an analytical lower bound for
\eqref{mutual-information} when the input is Gaussian distributed
to understand its performance in a simplistic manner.

\subsection{Lower Bound on Mutual Information}
We have the following result.\\

\textit{Proposition 4.5.1:}\,\,The mutual information of an iid
non coherent Rayleigh fading channel when the input distribution
is complex Gaussian, is lower bounded by
\begin{equation}\label{mi-lbfinal}
    I(X;Y)\geq \frac{1}{2}(C_{\text{cnf}}-C_{\text{rcsi}})
\end{equation} where $C_{\text{cnf}}$ and $C_{\text{rcsi}}$ are
the capacity of the non fading complex Gaussian channel and the
capacity of the Rayleigh fading channel when the CSI is perfectly
known at the receiver. The equality holds when the average input
power is zero.\\

\textit{Proof}: We consider $I(X;Y)$, $h(Y)$, and $h(Y|X)$ when
the input power $\sx^2$ is zero. Using \eqref{hyx-final}, we get
$h(Y|X)_{\sx^2=0}=-{\log{2}}+({1+\frac{\gamma}{2}})$. Since the
mutual information is zero with no channel input, we can write
\begin{equation}\label{equal-at-zero}
    h(Y)_{\sx^2=0}=h(Y|X)_{\sx^2=0}.
\end{equation}
The quantity $h(Y)$ in \eqref{out-entro} is monotonically
increasing with SNR, thus it has the minimum
\begin{equation}\label{hy-min}
    h(Y)_{\text{min}}=-{\log{2}}+({1+\frac{\gamma}{2}}).
\end{equation}
Consider, a non fading channel whose capacity achieving
distribution is Gaussian, where
$h(Y|X)_{\text{nf}}=h(N)=\frac{1}{2}{\log({\pi}{e}{\sn^2})}$ is
constant over input power. The monotonic increase of
$h(Y)_{\text{nf}}=\frac{1}{2}{\log[{\pi}{e}{(\sn^2+\sx^2)}]}$ with
SNR results in significant increase in channel capacity.\\
\begin{figure}
  \centering
  \includegraphics[width=\pictwidth]{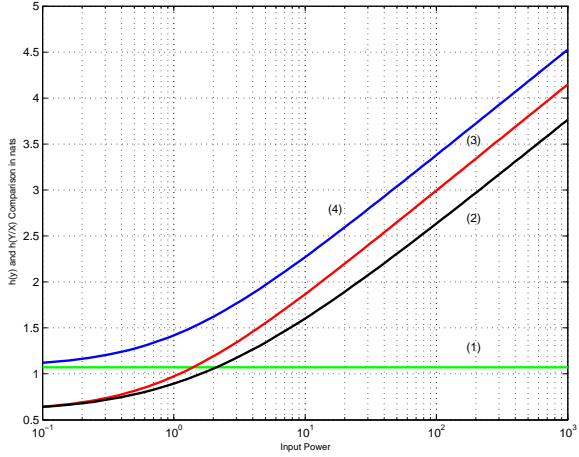}\\
  \caption{Entropy Comparison with Fading and Non fading Models:
(1) Output conditional entropy of non fading channel,
$h_{\text{nf}}(Y|X)$. (2) Output conditional entropy of fading
channel with Gaussian input. (3) Output entropy of fading channel,
$h(Y)$. (4) Output entropy of non fading channel,
$h(Y)_\text{{nf}}$. The $h(Y)_\text{{nf}}$ and $h(Y)$ treated in
this paper have a similar shape and the difference is decreasing
with SNR.}\label{figtwo}
\end{figure}

The fading introduces the monotonic increase of output conditional
entropy with SNR. In our investigation, we compare the
differential output entropies in both cases and use the properties
of the mutual information such as monotonic and non decreasing in
order to draw a lower bound. We consider a single dimension of the
non fading channel where the input and noise are complex random
variables and hence both the output and the conditional entropies
are taken accordingly. Fig. 2 portrays $h(Y)$ and $h(Y|X)$ of the
two channel models, where
\begin{equation}\label{hy-shannon}
    h(Y)_{\text{nf}}=\frac{1}{2}{\log[{\pi}{e}{(1+\sx^2)}]}
\end{equation}and
\begin{equation}\label{hyx-shannon}
    h(Y|X)_{\text{nf}}=\frac{1}{2}{\log({\pi}{e})}.
\end{equation}Note that
the abbreviation ``nf'' refers the Gaussian channel with no fading
present. Since the Gaussian distributions are the entropy
maximisers for a power limited input,
\begin{equation}\label{first-ineq}
    h(Y)_{\text{nf}}>h(Y) \,\,\forall\,\,\sx^2.
\end{equation}
Lets define the difference in \eqref{first-ineq}
\begin{equation}\label{the-gap-ver3}
    \gap=h(Y)_{\text{nf}}-h(Y),
\end{equation}
and investigate the bounds when $\sx^2=0$ and
$\sx^2\rightarrow\infty$. The $\gap_{\sx^2=0}$ can be written as
\begin{align}\label{max-gap}
\gap_{\sx^2=0}&=h(Y)_{nf,\sx^2=0}-h(Y)_{\text{min}}\nonumber\\
&=\frac{1}{2}{\log({\pi}{e})}+{\log{2}}-({1+\frac{\gamma}{2}}).
\end{align}
To calculate the difference when $\sx^2\rightarrow\infty$, we will
use the upper bound
\begin{equation}\label{shami-ixy}
    {\lim \atop {\sx^2\to \infty}}{I(X;Y)}\leq{\gamma}
\end{equation} given in \cite{Lapidoth-Shamai-2002}.
Using \eqref{hyx-final2} and \eqref{out-entro}, we can write the
mutual information of the channel as
\begin{equation}\label{ixy-final}
    I(X;Y)=h(Y)-\frac{1}{2}{C_{\text{rcsi}}}+{\log{2}}-({1+\frac{\gamma}{2}}).
\end{equation}

Substituting \eqref{shami-ixy} and \eqref{ixy-final} in
\eqref{the-gap-ver3}, we get,
\begin{align}\label{gap-infinity2}
    \gap_{\sx^2\rightarrow\infty}&\geq\frac{1}{2}\,{\lim \atop
    {\sx^2\to\infty}}\left\{{\log[{\pi}{e}{(1+\sx^2)}]}\right.\nonumber\\
    & \left. +\,{\exp\left({\frac{1}{\sx^2}}\right)}{\Ei\left(\frac{-1}{\sx^2}\right)}
    \right\}-\gamma+{\log{2}}-({1+\frac{\gamma}{2}})\nonumber\\
    &=\deltaa-\gamma+{\log{2}}-({1+\frac{\gamma}{2}}),
\end{align}
where $\deltaa=\frac{1}{2}[\gamma+\log({\pi}{e})]$. Refer the
Appendix B for the detailed proof. Therefore we can write
\eqref{gap-infinity2} as,
\begin{equation}\label{gap-infinity3}
\gap_{\sx^2\rightarrow\infty}\geq\log({2}{\sqrt{{\pi}{e}}})-(1+\gamma).
\end{equation}
Note that $\gap_{\sx^2=0}>\gap_{\sx^2\rightarrow\infty}$. The
differential entropies defined in here are monotonic and concave
with $\sx^2$. The gap defined in \eqref{the-gap-ver3} is the
difference between two monotonic concave functions which would not
necessarily be monotonic and concave. However, since
$\gap_{\sx^2=0}$ is higher than $\gap_{\sx^2\rightarrow\infty}$,
the quantity $G$ for any $\sx^2$ should be less than
$\gap_{\sx^2=0}$ due to the properties of the two entropies
mentioned. Therefore we conclude that the maximum difference
occurs at $\sx^2=0$. This $\gap_{\text{max}}=\gap_{\sx^2=0}$ can
be used to lower bound $h(Y)$ in \eqref{out-entro} and we get
\begin{equation}\label{hy-lbone}
    h(Y)\geq h(Y)_{\text{nf}}-\gap_{\text{max}}.
\end{equation}
Therefore, the mutual information in \eqref{ixy-final} can be
lower bounded as \begin{align}\label{mi-lbone}
    I(X;Y)&\geq h(Y)_{\text{nf}}-\gap_{\text{max}}-\left[\frac{1}{2}{C_{\text{rcsi}}}
    -{\log{2}}+({1+\frac{\gamma}{2}})\right]\nonumber\\
    &=h(Y)_{\text{nf}}-\frac{1}{2}{\log({\pi}{e})}-\frac{1}{2}{C_{\text{rcsi}}}\nonumber\\
    &=\frac{1}{2}{\log{(1+\sx^2)}}-\frac{1}{2}{C_{\text{rcsi}}},
\end{align}using \eqref{hyx-final2}, \eqref{hy-shannon}, and
\eqref{max-gap}. With $C_{\text{cnf}}={\log{(1+\sx^2)}}$, we
prove \eqref{mi-lbfinal}.\\
It should be noted here that \eqref{mi-lbfinal} asymptotically
converges to ${\gamma}/{2}$ since
${{\lim}\atop{{\sx^2}_\to\infty}}(C_{\text{cnf}}-C_{\text{rcsi}})={\gamma}$
\cite{Lapidoth-Shamai-2002}.

\begin{figure}
  \centering
  \includegraphics[width=\pictwidth]{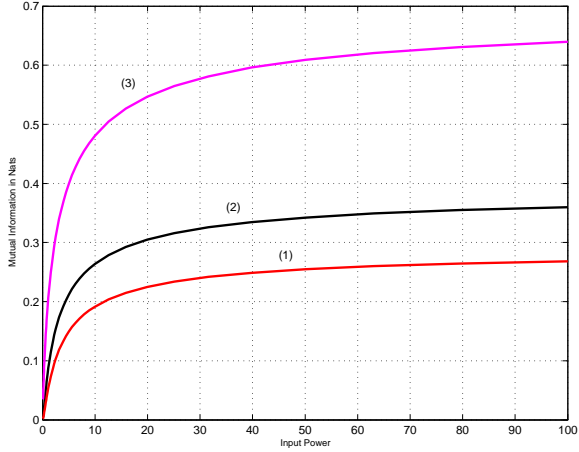}\\
  \caption{Analysis of the lower bound in fading channel:
  (1) The lower bound with Gaussian input. (2) The mutual information acquired using the closed form expression with
  Gauss-Hermit quadrature. (3) The channel capacity achieved with a discrete input.
   }\label{figthree}
\end{figure}

\begin{figure}
  \centering
  \includegraphics[width=\pictwidth]{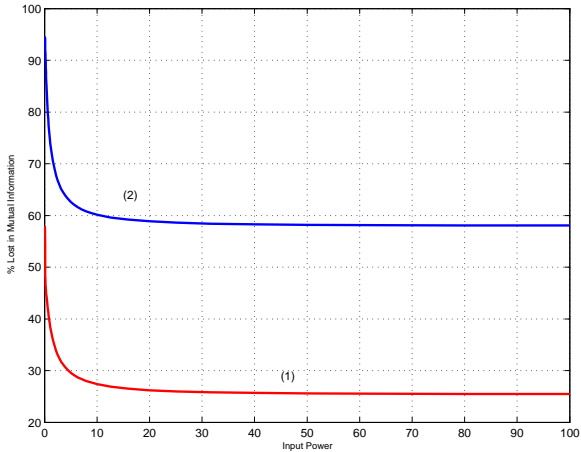}\\
  \caption{Percentage lost using the lower bound: (1) With the mutual information found numerically
   for Gaussian input. (2) With the channel capacity.}\label{figfour}
\end{figure}

\section{NUMERICAL RESULTS}
We compare the new lower bound with the mutual information found
with the closed form expression attained through Gauss-Hermit
quadrature in the previous section.

The lower bound in \eqref{mi-lbfinal} is plotted against the input
power in Fig. 3 with the mutual information obtained using the
closed form expression. Further, it is compared with channel
capacity acquired with discrete input
\cite{DTMRayleigh-Shami-2001}. The channel capacity is plotted for
comparison only with two discrete mass points one located at the
origin since the probability of other mass points are small at low
SNR and even suited for a simple comparison at high SNR due to the
percentage increase in capacity is low \cite{Abou-1996}. The
percentage lost in mutual information with a Gaussian input on our
lower bound is plotted in Fig. 4 where it shows $30\%$ less than
the numerical values.

%
%

\section{CONCLUSIONS}
The mutual information of a non-coherent Rayleigh fading channel
when the input is Gaussian distributed can be expressed in closed
form using Gauss-Hermit quadrature. Further it can be lower
bounded as the difference between the capacities of non fading
channel and the Rayleigh fading channel when the perfect channel
state information is known at the receiver. Even the Gaussian
input is not optimal, our result shows the minimum achievable
information rate which can be used as the worse case scenario in
non coherent Rayleigh fading channels. The lower bound found is
never lower than $70\%$ of the actual.

The CSI is obtained by training with known pilot symbols inserted
in the transmitted sequence. Due to the presence of noise or under
the fast fading conditions, the receiver is provided with
imperfect CSI and the performance of the channel depends on its
quality. Considering the worst case scenario, the channel can
become non coherent with Gaussian input which optimises the mutual
information with perfect CSI. Therefore, the closed form
expression shown in this paper could be used as the lower bound
with the imperfect CSI at the receiver.

\section{APPENDIX}
\subsection{PROOF OF CONDITIONAL ENTROPY IN \eqref{hyx-final}}
We write \eqref{hyx-one} as \begin{equation}\label{app-hyx}
   h(Y|X)=E_1-{\log{2}}+({1+\frac{\gamma}{2}})
\end{equation} where \begin{equation}\label{appE1-integral}
    E_1={{\lim}\atop{{k_1}_\to\infty}}{\int_{0}^{k_1}}
    {\frac{{x}}{k_2}}{\exp\left({\frac{-x^2}{k_2}}\right)}{\log(1+x^2)}{dx}, \,\,x\geq0
\end{equation} and $k_2=\sx^2$.

Consider the integral part of \eqref{appE1-integral}. Using
integration by parts, we get
\begin{multline}
    \int_{0}^{k_1}\frac{x}{k_2}{\exp\left({\frac{-x^2}{k_2}}\right)}{\log(1+x^2)}{dx}=\\
    \left\{-\frac{1}{2}\exp\left(\frac{-x^2}{k_2}\right){\log{(1+x^2)}}\right\}_{0}^{k_1}\\
    +\int_{0}^{k_1}{\exp\left({\frac{-x^2}{k_2}}\right)}\frac{x}{(1+x^2)}{dx}.
\label{moi-intparts}
\end{multline}
Substituting $t=1+x^2$, the
second term of \eqref{moi-intparts} can be written as
\begin{align}\label{moi-intparts2}
 \int_{0}^{k_1}{\exp\left({\frac{-x^2}{k_2}}\right)}\frac{x\,\,{dx}}{(1+x^2)}=\frac{1}{2}{\exp\left(\frac{1}{k_2}\right)}\int_{1}^{1+{k_1}^2}
  \frac{{\exp\left(\frac{-t}{k_2}\right)}}{t}{dt}.
\end{align} Substituting $u=t/k_2$ in the right hand side of \eqref{moi-intparts2} we get
\begin{align}\label{moi-intparts3}
    \int_{0}^{k_1}{\exp\left({\frac{-x^2}{k_2}}\right)}\frac{x\,\,{dx}}{(1+x^2)}&=\frac{1}{2}{\exp\left(\frac{1}{k_2}\right)}\int_{\frac{1}{k_2}}^{\frac{1+{k_1}^2}{k_2}}
  \frac{e^{-u}}{u}{du}\nonumber\\
  &=\frac{1}{2}{\exp\left(\frac{1}{k_2}\right)}\left(\int_{\frac{1}{k_2}}^{\infty}\frac{e^{-u}}{u}{du}\right.\nonumber\\
  & \left.\,-\int_{\frac{1+{k_1}^2}{k_2}}^{\infty}\frac{e^{-u}}{u}{du}\right)\nonumber\\
  &=\frac{1}{2}{\exp\left(\frac{1}{k_2}\right)}\left[\Ei\left(-\frac{1+x^2}{k_2}\right)\right]_{0}^{k_1}.
\end{align}
Using this identity in \eqref{moi-intparts} we get
\begin{multline}\label{moi-intparts4}
 \int_{0}^{k_1}\frac{x}{k_2}{\exp\left({\frac{-x^2}{k_2}}\right)}{\log(1+x^2)}{dx}=\\
 \frac{1}{2}{\exp\left(\frac{1}{k_2}\right)}\left[\Ei\left(-\frac{1+x^2}{k_2}\right)\right]_{0}^{k_1}\\
 -\left[\frac{1}{2}\exp\left(\frac{-x^2}{k_2}\right){\log{(1+x^2)}}\right]_{0}^{k_1}.
\end{multline} Now we can write \eqref{appE1-integral} as
\begin{align}\label{appE1simp}
    E1&={{\lim}\atop{{k_1}_\to\infty}}\frac{1}{2}\left[\exp\left({\frac{1}{k_2}}\right)\Ei\left(-\frac{1+{k_1}^2}{k_2}\right)\right.\nonumber\\
    & \left.\,-\exp\left(\frac{-{k_1}^2}{k_2}\right){\log{(1+{k_1}^2)}}\right]-\frac{1}{2}\exp\left({\frac{1}{k_2}}\right)\Ei\left(-\frac{1}{k_2}\right).
\end{align} By applying La'Hospital's Rule, it can be shown that \begin{equation}\label{Lahopiapp}
    {{\lim}\atop{{k_1}_\to\infty}}\frac{1}{2}\exp\left(\frac{-{k_1}^2}{k_2}\right){\log{(1+{k_1}^2)}}=0.
\end{equation} Also note that $\Ei{(-\infty)}=0$
\cite{Hand-book-mathematical-functions}, thus
\begin{equation}\label{E1final}
    E_1=-\frac{1}{2}\exp\left({\frac{1}{k_2}}\right)
    \Ei\left(-\frac{1}{k_2}\right).
\end{equation} By substituting \eqref{E1final} in \eqref{app-hyx}
completes the proof.

\subsection{PROOF OF THE ASYMPTOTIC ANALYSIS USED IN \eqref{gap-infinity2}}
Let's define $\power=\sx^2$ and we write the asymptotic value in
\eqref{gap-infinity2} as,
\begin{equation}\label{l-value1}
    \deltaa=\frac{1}{2}{\lim \atop
{\power\to\infty}}\left[{\log[{\pi}{e}{(1+\power)}]}+{\exp\left({\frac{1}{\power}}\right)}{\Ei\left(\frac{-1}{\power}\right)}\right]
\end{equation}
where the exponential integral can be expressed as,
\cite{Table-of-integrals}
\begin{equation}\label{ei-defined}
    \Ei(-x)=\gamma+{e^{-x}}{\log{x}}+\int_{0}^{x}{e^{-t}{\log{t}{dt}}}.
\end{equation} Using this identity we get,

\begin{align}
\deltaa&=\frac{1}{2}{\lim \atop
{\power\to\infty}}\left\{{\log[{\pi}{e}{(1+\power)}]}\right.\nonumber\\
&\left.\,+{\exp\left({\frac{1}{\power}}\right)}\left[\gamma+{\exp\left({\frac{-1}{\power}}\right){\log{\frac{1}{\power}}}}\right]\right\}\nonumber\\
&+\frac{1}{2}{\lim \atop
{\power\to\infty}}{\exp\left({\frac{1}{\power}}\right)}\left(\int_{0}^{\exp\left({\frac{1}{\power}}\right)}{e^{-t}{\log{t}{dt}}}\right)\nonumber\\
&=\frac{1}{2}{\lim \atop
{\power\to\infty}}\left\{{\log[{\pi}{e}{(1+\frac{1}{\power})}]}+{\gamma}{\exp\left({\frac{1}{\power}}\right)}\right\}+0\nonumber\\
&=\frac{1}{2}[{\gamma}+{\log({\pi}{e})}]
\end{align}
which competes the proof.


\section{ACKNOWLEDGEMENTS}
National ICT Australia (NICTA) is funded through the Australian
Government's \emph{Backing Australia's Ability Initiative}, in
part through the Australian Research Council.


\end{document}